\documentclass[useAMS,usenatbib]{mn2e}

\usepackage{amsmath}
\usepackage{comment}
\usepackage{graphicx}
\usepackage{rotating}
\usepackage{color}
\usepackage{aas_macros}

\newcommand{\beq}{\begin{equation}}
\newcommand{\enq}{\end{equation}}
\newcommand{\beqa}{\begin{eqnarray}}
\newcommand{\enqa}{\end{eqnarray}}
\newcommand{\beit}{\begin{itemize}}
\newcommand{\enit}{\end{itemize}}
\newcommand{\bem}{\begin{pmatrix}}
\newcommand{\enm}{\end{pmatrix}}

\newcommand{\vecalpha}{\boldsymbol{\alpha}}	

\newcommand{\vecx}{\mathbf{x} }
\newcommand{\vecy}{\mathbf{y} }

\newcommand{\veck}{\mathbf{k} }
\newcommand{\vecv}{\mathbf v}

\newcommand{\vecu}{\mathbf u}

\newcommand{\Tr}{\mathrm{Tr}}

\newcommand{\lat}{\left\langle}
\newcommand{\rat}{\right\rangle}
\newcommand{\av}[1]{\lat #1 \rat}
\newcommand{\Var}[1]{\textrm{Var}\left(#1\right)}

\newcommand{\obs}{\textrm{obs}}

\newcommand{\cst}{\textrm{cst}}
\newcommand{\eff}{\textrm{eff}}
\newcommand{\lb}{\left [}
\newcommand{\rb}{\right ]}
\newcommand{\lp}{\left (}
\newcommand{\rp}{\right )}

\renewcommand{\max}{\mathrm{max}}
\renewcommand{\min}{\mathrm{min}}
\newcommand{\inv}{^{-1}}

\newcommand{\Fab}{F_{\alpha\beta}}
\newcommand{\bes}{\begin{sideways}}
\newcommand{\ees}{\end{sideways}}

\renewcommand{\bbeta}{\boldsymbol {\eta}}
\renewcommand{\bmu}{\boldsymbol {\mu}}
\newcommand{\vecl}{\mathbf l}

\newcommand{\black}{\color{black}}

\title[Probe combination in large galaxy surveys]{Probe combination in large galaxy surveys : \\ Application of Fisher information and Shannon entropy to weak lensing}
\author[Carron et al]{J. Carron\thanks{E-mail:
jcarron@phys.ethz.ch}, A. Amara, and S.J. Lilly  \\
Insitute of Astronomy, ETH Zuerich, Switzerland}
\begin{document}

\date{Accepted 2011 July 4.  Received 2011 June 21; in original form 2010 December 6}

\pagerange{\pageref{firstpage}--\pageref{lastpage}} 

\maketitle

\label{firstpage}

\begin{abstract}
This paper is aimed at developing a better understanding of the structure of the information that in contained in galaxy surveys, so as  to find optimal ways to combine observables from such surveys. We first show how Jaynes' Maximal Entropy Principle allows us, in the general case, to express the Fisher information content of data sets in terms of the curvature of the Shannon entropy surface with respect to the relevant observables. This allows us to understand the Fisher information content of a data set, once a physical model is specified, independently of the specific way that the data will be processed, and without any assumptions of Gaussianity. This includes as a special case the standard Fisher matrix prescriptions for Gaussian variables widely used in the cosmological community, for instance for power spectra extraction.
As an application of this approach, we evaluate the prospects of a joint analysis of weak lensing tracers up to second order in the shapes distortions, in the case that the noise in each probe can be effectively treated as model independent. These include the magnification, the two ellipticity and the four flexion fields. At the two point level, we show that the only effect of treating these observables in combination is a simple scale dependent decrease of the noise contaminating the accessible spectrum of the lensing E-mode. We provide simple bounds to its extraction by a combination of such probes, as well as its quantitative evaluation when the correlations between the noise variables for any two such probes can be neglected.
\end{abstract}


\section{Introduction} 

With cosmological data sets currently going through a rapid period of growth, it is increasingly important to quantitatively understand the potential and limits of particular data sets to test a physical model or hypothesis. For this, the Fisher information matrix has become a widely used tool in cosmology.

The concept of Fisher information has a long history. It was first coined by the statistician and geneticist R.A. Fisher \citep{fisher25} under the name of intrinsic accuracy of frequency curves. It has found its way into the cosmological community over the last decade, where it is often used to optimise survey configurations \citep{2007MNRAS.381.1018A,2007MNRAS.377..185P,2006astro.ph..9591A,2009ApJ...695..652B} of planned cosmology experiments or to evaluate the expected errors on certain cosmological parameters with some observables \citep{1997ApJ...480...22T,Tegmark97,1999ApJ...514L..65H,2004PhRvD..70d3009H}. 

Much of the work to date has been limited to particular sets of observables and estimators. Usually, it is assumed that observational errors as well as the parameters probability distribution have Gaussian shape. The first aim of this work is to propose a framework to express the global Fisher information content of large data sets in a way that is independent of the specific ways that the data will be processed, and in realistic situations, where the exact statistical properties of the data are not known precisely. This should then provide a well motivated basis point in order to perform systematic and robust trade-off studies. For this purpose a number of useful concepts already exist in  the fields of information theory and probability theory, such as Shannon entropy or relative entropy \citep{Kullback}, which we can use to gain a better understanding of what we can achieve with planned experiments. Specifically, we will show that we can achieve our aim by combining Fisher's information measure with Jaynes' Principle of Maximal Entropy.  \citep{jaynes83,jaynes2003}.
\newline\indent 
In a second step, as a concrete application of this approach, we investigate the joint entropy and information content of multiple observables of the same underlying, cosmologically interesting field. This is a very relevant situation in weak lensing \citep{1992grle.book.....S,2001PhR...340..291B,2003ARA&A..41..645R,2006astro.ph.12667M,2006glsw.book.....S}, where the distortions of galaxy images to any order are sourced by the lensing potential field. \newline\newline\indent
This paper is divided into the following sections. In section \ref{Setting the scene},  we present in details our approach.  We first review and develop some key properties of Fisher information, and its link to the Cramer Rao inequality. We put a strong emphasis on its interpretation as a measure of information on model parameters in a data set, that is obtained from the probability distribution of different observational outcomes as function of these same model parameters. Readers familiar with these aspects may jump to section {\ref{The structure of the information in a physical model}}, where we introduce Jaynes' Maximum Entropy Principle, and show how it ideally completes Fisher's information measure, allowing us to understand the information content of a data set on a physical model in the case of incomplete knowledge.
In section \ref{application}, we show how the study of the Shannon entropy of a set of homogeneous fields provide a simple and model parameter independent answer to the question of the combination of the weak lensing observables shear, magnification and flexion.
We provide in section \ref{results}  quantitative evaluation of the prospects of such a combination at the two-point level for typical dark energy surveys parameters, and conclude in section \ref{discussion} with a summary of the results and a discussion. A set of appendices collects some technical details.
\section{Fisher Information and Jaynes MaxEnt principle}
\label{Setting the scene}
The concept of Fisher information is rich and not limited to parameter error estimation. We review here a few simple points of interest that justify the interpretation of the Fisher matrix as a measure of the information content of an experiment. Let us begin by considering the case of a single measurement $X$, with different possible outcomes, or realisations, $x$, and our model has a single parameter $\alpha$. We also assume that we have knowledge, prior to the given experiment, of the probability density function $p_X(x,\alpha)$, which depends on our parameter $\alpha$, that gives the probability of observing particular realisations for each value of the model parameter.
The Fisher information, $F$, in $X$ on $\alpha$, is a non-negative scalar in this one parameter case. It is defined  in a fully general way as a sum over all realisations of the data \citep{fisher25}:
\beq \label{FisherI}
F^X(\alpha) = \av{\left( \frac{\partial \ln p_X(x,\alpha)}{\partial\alpha} \right)^2}.
\enq
Angle brackets will always stand for mean value with respect to the probability density function, i.e. for any function $f$,
\beq
\av{f} \equiv \int dx\: p_X(x,\alpha) f(x).
\enq
Three simple but important properties of Fisher information are worth highlighting at this point. \newline

\indent The first is that $F^X(\alpha)$ is positive definite, and it vanishes if and only if the parameter $\alpha$ does not impact the data, i.e. if the derivative of $p_X(x,\alpha)$ with respect to $\alpha$ is zero for every realisation $x$.
\newline

\indent The second point is that it is invariant to invertible manipulations of the observed data. This can be seen by considering an invertible change of variable $y = f(x)$, which, due to the rules of probability theory can be expressed as 
\beq
p_Y(y,\alpha) = p_X(x,\alpha)\left |\frac {dx}{dy}\right|.
\enq
Thus
\beq
\frac{\partial \ln p_Y(y,\alpha)}{\partial \alpha} = \frac{\partial \ln p_X(x,\alpha)}{\partial \alpha},
\enq 
leading to the simple equivalence that $F^X(\alpha) = F^Y(\alpha)$. On the other hand, information may be lost when the transformation is not unique in both directions. \citep[e.g., see][for a proof]{Rao}. For instance, if the data is combined to produce a new variable that could arise from different sets of data points.
This is only the statement that manipulations of the data leads, at best, only to conservation of the information.\newline

\indent The third point is that information from independent experiments add together. Indeed, if two experiments with data $X$ and $Y$ are independent, then the joint probability density factorises,
\beq
p_{XY}(x,y) = p_X(x)p_Y(y),
\enq
and it is easy to show that the joint information in the observations decouples,
\beq
F^{XY}(\alpha) = F^X(\alpha) + F^Y(\alpha).
\enq\newline
\indent 
These properties are making the Fisher information a meaningful measure of information. This is independent of its interpretation as providing error bars on parameters. It further implies that once a physical model is specified with a given set of parameters, a given experiment has a definite information content that can only decrease with data processing.
\subsection{The case of a single observable}\label{single observable}
To quantify the last point above, and in order to get an understanding of the structure of the information in a data set, we first review a simple situation, common in cosmology, where the extraction of the model parameter $\alpha$ from the data goes through the intermediate step of estimating a particular observable, $D$, from the data, $x$, with the help of which $\alpha$  will be inferred. A typical example could be, from the temperature map of the  CMB ($x$), the measurement of the power spectra of the fluctuations ($D$), from which a cosmological parameter ($\alpha$) is extracted.
The observable $D$ is measured from $x$ with the help of an estimator, that we call $\hat D$, and that we will take as unbiased. This means that its mean value, as would be obtained for instance if many realizations of the data were available, converges to the actual value that we want to compare with the model prediction,
\beq
\lat \hat D\rat = D(\alpha).
\enq
A measure for its deviations from sample to sample, or the uncertainty in the actual measurement, is then given by the variance of $\hat D$, defined as
\beq \label{variance}
\Var{\hat D} = \av{\hat D^2} - \av{\hat D}^2.
\enq
In such a situation, a major role is played by the so-called Cram\'{e}r-Rao 
 inequality (\cite{Rao}),  that links the Fisher information content of the data to the variance of the estimator, stating that
\beq \label{CR}
\textrm{Var}(\hat D) F^X(\alpha) \ge \left(\frac{\partial D(\alpha)}{\partial \alpha}\right)^2.
\enq
This equation holds for any such estimator $\hat D$ and any model parameter $\alpha$. Two different interpretations of this equation are possible: \newline

\indent The first bounds the variance of $\hat D$ by the inverse of the Fisher information. To see this, we consider the special case of the model parameter $\alpha$ being $D$ itself. Although we are making in general a conceptual distinction between the observable $D$ and the model parameter $\alpha$, nothing requires us from doing so. Since $\alpha$ is now equal to $D$, the derivative on the right hand side becomes unity, and one obtains
\beq \label{CRB}
\textrm{Var}(\hat D) \ge \frac 1 {F^X(D)}.
\enq
The variance of any unbiased estimator $\hat D$ of  $D$ is therefore bounded by the inverse of the amount of information $F^X(D)$ the data possess on $D$. If $F^X(D)$ is known it gives a useful lower limit on the error bars that the analysis of the data can put on this observable.
\newline
\newline
\indent The second reading of the Cram\'{e}r-Rao 
 inequality, closer in spirit to the present work, is to look at how information is lost by constructing the observable $D$, and discarding the rest of the data set. For this, we rewrite trivially equation (\ref{CR}) as
\beq \label{CramerRao}
F^X(\alpha) \ge  \left(\frac{\partial D}{\partial \alpha}\right)^{2} \frac 1 {\textrm{Var}(\hat D)}.
\enq
The expression on the right hand side is the ratio of the sensitivity of the observable to the model parameter $\lp \frac{ \partial D}{\partial \alpha}\rp^2 $, to the accuracy with which the observable can be extracted from the data, $\textrm{Var}(\hat D)$. One of the conceivable approaches in order to estimate the true value of the parameter $\alpha$, is to perform a $\chi^2$ fit to the measured value of $D$ . It is simple to show that this ratio, evaluated at the best fit value, is in fact proportional to the expected value of the curvature of $\chi^2(\alpha)$ at this value. Since the curvature of the $\chi^2$ surface describes how fast the value of the $\chi^2$ is increasing when moving away from the best fit value, its inverse can be interpreted as an approximation to the error bar that the analysis with the help of  $\hat D$ will put on $\alpha$. 
\newline
\newline
Thus, equation (\ref{CramerRao}) shows that by only considering $D$ and not the full data set, we may have lost information on $\alpha$, a loss given by the difference between the left and right hand side of that equation. While the latter may be interpreted as the information on $\alpha$ contained in the part of the data represented by $D$, we may have lost trace of any other source of information. 
\subsection{The general case} \label{CramerRaoMulti}
These considerations on the Cram\'{e}r-Rao 
 bound can be easily generalised to the case of many parameters and many estimators of as many observables. Still dealing with a measurement $X$ with outcomes $x$, we want to estimate a set of parameters
\beq
\boldsymbol \theta = (\alpha, \beta,\cdots)
\enq
 with the help of some vector of observables,
\beq
\mathbf D = (D_1,\cdots,D_n)
\enq
that are extracted from $x$ with the help of an array of unbiased estimators,
\beq
\mathbf {\hat D} = \bem \hat D_1,\cdots, \hat D_n \enm,\quad \av{\mathbf{\hat D}} = \mathbf D
\enq
In this multidimensional setting, all the three scalar quantities that played a role in our discussion in section \ref{single observable}, i.e. the variance of the estimator, the derivative of the observable with respect to the parameter, and the Fisher information, are now matrices.
\newline
\newline
The Fisher information $F$ in $X$ on the parameters $\boldsymbol \theta$ is defined as the square matrix
\beq \label{FisherII}
\lb F^X\lp \boldsymbol \theta \rp \rb_{\alpha\beta} = \av{ \ \frac {\partial \ln p_X}{\partial \alpha}\frac{\partial\ln p_X}{\partial \beta}}.
\enq
While the diagonal elements are identical to the information scalars in equation (\ref{FisherI}), the off diagonal ones describe correlated information.
The Fisher information matrix still carries the three properties we discussed in section \ref{Setting the scene}.
\newline
The variance of the estimator in equation (\ref{variance}) now becomes the covariance matrix $\textrm{cov}(\mathbf {\hat D})$ of the estimators $\mathbf {\hat D}$, defined as
\beq
\textrm{cov} \lp \mathbf {\hat D} \rp _{ij} = \av{\hat D_i \hat D_j} - D_iD_j.
\enq
Finally, the derivative of the observable with respect to the parameter, in the right hand side of (\ref{CR}), becomes a matrix $\Delta$, in general rectangular, defined as
\beq
\Delta_{\alpha \:i} = \frac{\partial D_i}{\partial \alpha},
\enq
where $\alpha$ runs over all elements of the set $\btheta$ of model parameters.
Again, the Cram\'{e}r-Rao 
 inequality provides a useful link between these three matrices, and again there are two approaches to that equation : first, as usually presented in the literature \citep{Rao}, in the form of a lower bound to the covariance matrix of the estimators,
\beq \label{CRmulti}
\textrm{cov} \left( \mathbf {\hat D} \right) \ge \Delta^T \lb  F^X\lp \boldsymbol \theta \rp \rb ^{-1}  \Delta.
\enq
The inequality between two symmetric matrices $A \ge B$ having the meaning that the matrix $A - B$ is positive definite \footnote{A matrix $A$ is called positive definite when for any vector $x$ holds that $x^TAx \ge 0$.  A concrete implication for our purposes is e.g. that the diagonal entries of the left hand side of (\ref{CRmulti}) or (\ref{covF}), which are the individual variances of each estimator $\hat D_i$, are greater than those of the right hand side.   For many more properties of positive definite matrices, see for instance \citep{Bhatia}}. If, as above, we consider the special case of identifying the parameters with the observables themselves, the matrix $\Delta$ is the identity matrix, and so we obtain that the covariance of the vector of the estimators is bounded by the inverse of the amount of Fisher information that there is on the observables in the data,
\beq \label{covF}
\textrm{cov}(\mathbf {\hat D}) \ge \lb F^X(\mathbf D)\rb^{-1}.
\enq
\newline
 Second, we can turn this lower bound on the covariance to a lower bound on the amount of information in the data set as well. By rearranging equation (\ref{CRmulti}), we obtain the multidimensional analogue of equation (\ref{CramerRao}), which describes the loss of information that occurs when the data is reduced to a set of estimators,
\beq \label{MultiCramer}
 F^X\lp \boldsymbol \theta \rp \ge \Delta \lb \textrm {cov} \lp \mathbf {\hat D }\rp\rb^{-1} \Delta^{T}.
\enq
For the sake of completeness, a proof of these two inequalities can be found in Appendix \ref{CramerRao bound}.
\newline
\newline
Instead of giving a useful lower bound to the covariance of the estimator as in equation (\ref{CRmulti}), in this form the Cram\'{e}r-Rao 
 inequality makes clear how information is in general lost when reducing the data to any particular set of estimators. The right hand side may be seen, as before, as the expected curvature of a $\chi^2$ fit to the estimates produced by the estimators $\mathbf {\hat D}$, when evaluated at the best fit value, with all correlations fully and consistently taken into account.\newline 

\noindent In the next two sections, we show how Jaynes' Maximal Entropy Principle allow us to understand the total information content of a data set, once a model is specified, in very similar terms.
\subsection{Jaynes Maximal Entropy Principle}
\label{The structure of the information in a physical model}
In cosmology,  the knowledge of the probability distribution  of the data as function of the parameters, $p_X(x,\btheta)$,  which is compulsory in order to evaluate its Fisher information content, is usually very limited. In a galaxy survey, a data outcome $x$ would be typically the full set of angular positions of the galaxies, together with some redshift estimation if available, to which we may add any other kind of information, such as luminosities, shapes, etc.  Our ignorance of both initial conditions and of many relevant physical processes does not allow us to predict either galaxy positions in the sky, or all interconnections with all this additional information.  Our predictions of the shape of $p_X$ is thus limited to some statistical properties, that are sensitive to the model parameters $\btheta$, such as the mean density over some large volume, or certain types of correlation functions.\newline\newline
In fact, even if it were possible to devise some procedure in order to get the exact form of $p_X$, it may eventually turn out to be useless, or even undesirable, to do so. The incredibly large number of degrees of freedom of such a function is very likely to overwhelm the analyst with a mass of irrelevant details,  which may have no relevant significance on their own, or improve the analysis in any meaningful way.
\newline 
\newline
These arguments call for a kind a thermodynamical approach, which would try and capture those aspects of the data which are relevant to our purposes, reducing the number of degrees of freedom in a drastic way. Such an approach already exists in the field of probability theory \citep{jaynes57}. It is based on Shannon's concept of entropy of a probability distribution \citep{shannon48} and shed new light on the connection between probability theory and statistical mechanics.
\newline
\newline
As we have just argued, our predictive knowledge of $p_X(x,\btheta)$ is limited to some statistical properties. Let us formalise this mathematically, in a similar way as in section \ref{CramerRaoMulti}.  Astrophysical theory gives us a set of constraints on the shape of $p_X$, in the form of averages of some functions $o_i$,
\beq \label{constraints} 
O_i(\btheta) = \lat o_i(x) \rat (\btheta), \quad i = 1,\cdots,n.
\enq
where $p_X$ enters through the angle brackets. As an example, suppose the data outcome $x$ is a map of the matter density field as a function of position. In this case, one of these constraints  $O_i$  could be the mean of the field or its power spectrum, as given by some cosmological model.
\newline\newline
The role of this array $\mathbf O = (O_1,\cdots,O_n)$ is to represent faithfully the physical understanding we have of $p_X$, according to the model, as a function of the model parameters $\btheta$.  In the ideal case, some way can be devised to extract each one of these quantities $O_i$ from the data and to confront them to theory. 
\newline\newline
The set of observables $\mathbf D$, that we used in section \ref{CramerRaoMulti},  would be a subset of these predictions $\mathbf O$, and we henceforth refer to $\mathbf O $ as the 'constraints'.
\subsection{Maximal entropy distributions} 
Although $p_X$ must satisfy the constraints (\ref{constraints}), there may still be a very large number of different distributions compatible with these. However, a very special status among these distributions has the one which maximises the value of Shannon's entropy\footnote{Formally, for continuous distributions the reference to another distribution is needed to render S invariant with respect to invertible transformations, leading to the concept of the entropy of $p_X$ relative to another distribution $q_X$, $S = \int dx \:p_X(x)\ln \frac {p_X(x)}{q_X(x)}$, also called Kullback-Leibler divergence. The quantity defined in the text is more precisely the entropy of $p_X(x)$ relative to a uniform probability density function. For an recent account on this, close in spirit to this work, see \cite{caticha08}.}, defined as
\beq\label{Entropy}
S = - \int dx \:p_X(x,\btheta)\ln p_X(x,\btheta).
\enq
First introduced by Shannon \citep{shannon48} as a  measure of the uncertainty in a distribution on the actual outcome, Shannon's entropy is now the cornerstone of information theory. Jaynes' Maximal Entropy Principle states that the $p_X$  for which this measure $S$ is maximal is the one that best deals with our insufficient knowledge of the distribution, and should be therefore preferred. We refer the reader to Jaynes' work \citep{jaynes83,jaynes2003} and to \cite{caticha08} for detailed discussions of the role of entropy in probability theory and for the conceptual basis of maximal entropy methods. Astronomical applications related to some extent to Jaynes's ideas include image reconstruction from noisy data, (see e.g. \citep{1984MNRAS.211..111S,1996VA.....40..563S,2004MNRAS.347..339M} and references therein) , mass profiles reconstruction from shear estimates \citep{1998MNRAS.299..895B,2002MNRAS.335.1037M}, as well as model comparison when very few data is available \citep{2007MNRAS.380..865Z}. We will see that for our purposes as well it provides us a powerful tool, and that the Maximal Entropy Principle is the ideal complement to Fisher information, fitting very well within our discussions in section \ref{Setting the scene} on the Cram\'{e}r-Rao 
 inequality.\newline
\newline
Intuitively, the entropy $S$ of $p_X$ tells us how sharply constrained the possible outcomes $x$ are, and Jaynes' Maximal Entropy Principle selects the $p_X$ which is as wide as possible, but at the same time consistent with the constraints (\ref{constraints}) that we put on it. 
The actual maximal value attained by the entropy $S$, among all the possible distributions which satisfy (\ref{constraints}), is a function of the constraints $\mathbf O$, which we denote by
\beq \label{EntropyII}
S(O_1,\cdots,O_n).
\enq
Of course it is a function of the model parameters $\btheta$ as well, since they enter the constraints.  As we will see, the shape of that surface as a function of $\mathbf O$, and thus implicitly as a function of $\btheta$, is the key point in understanding the Fisher information content of the data. In the following, in order to keep the notation simple, we will omit the dependency on $\btheta$ of most of our expressions, though it will always be implicit.
\newline
\newline
The problem of finding the distribution $p_X$ that maximises the entropy (\ref{Entropy}), while satisfying the set of constraints (\ref{constraints}), is an optimization exercise. We can quote the end result \cite[chap. 11]{jaynes83},\cite[chap. 4]{caticha08}: \newline
The probability density function $p_X$, when it exists, has the following exponential form,
\beq \label{MaxEnt}
p_X(x) = \frac 1 {Z} \exp\left(-\sum_{i = 1}^{n}\lambda_i o_i(x)\right),
\enq
in which to each constraint $O_i$ is associated a conjugate quantity $\lambda_i$, that arises formally as a Lagrange multiplier in this optimization problem with constraints. The conjugate variables $\lambda$'s are also called  'potentials', terminology that we will adopt in the following. We will see below in equation (\ref{potentials}) that the potentials have a clear interpretation, in the sense that the each potential $\lambda_i$ quantifies how sensitive is the entropy function $S$ in (\ref{EntropyII}) to its associated constraint $O_i$. The quantity $Z$, that plays the role of the normalisation factor, is called the partition function. Since equation (\ref{MaxEnt}) must integrate to unity, the explicit form of the partition function is 
\beq \label{partitionfunction}
Z(\lambda_1,\cdots,\lambda_n) = \int dx \: \exp\left(-\sum_{i=1}^{n}\lambda_i o_i(x)\right).
\enq
The actual values of the potentials are set by the constraints (\ref{constraints}). They reduce namely, in terms of the partition function, to a system of equations to solve for the potentials,
\beq \label{lambda}
 O_i = -\frac{\partial}{\partial \lambda_i}\ln Z,\quad i = 1,\cdots,n.
\enq
The partition function $Z$ is closely related to the entropy $S$ of $p_X$. It is simple to show that the following relation holds,
\beq \label{legendre}
S = \ln Z + \sum_{i = 1}^n \lambda_iO_i,
\enq
and the values of the potentials can be explicitly written as function of the entropy, in a relation mirroring equation (\ref{lambda}),
\beq \label{potentials}
 \lambda_i =  \frac{\partial S}{\partial O_i}, \quad i = 1,\cdots,n
\enq
Given the nomenclature, it is of no surprise that a deep analogy between this formalism and statistical physics does exist. Just as the entropy, or partition function, of a physical system determines the physics of the system, the statistical properties of these maximal entropy distributions follow from the functional form of the Shannon entropy or its partition function as a function of the constraints. For instance, the covariance matrix of the constraints is given by
\beq \label{fluctuations}
\lat \left(o_i(x) - O_i\right)\left(o_j(x) - O_j\right)\rat = \frac{\partial^2\ln Z}{\partial \lambda_i\partial\lambda_j} 
\enq
In statistical physics the constraints can be the mean energy, the volume or the mean particle number, with potentials being the temperature, the pressure and the chemical potential. We refer to \cite{jaynes57} for the connection to the physical concept of entropy in thermodynamics and statistical physics.
\subsection{The structure of the information in large data sets} \label{combinations}
With our choice of probabilities $p_X$ given by equation (\ref{MaxEnt}), the amount of Fisher information on the parameters $\btheta = (\alpha,\beta,\cdots)$ of the model can be evaluated in a straightforward way. The dependence on the model goes through the constraints, or, equivalently, through their associated potentials. It holds therefore that
\beq \begin{split}
\frac{ \partial \ln p_X(x)}{\partial \alpha}& =-\frac {\partial \ln Z}{\partial \alpha} -\sum_{i= 1}^n\frac{\partial\lambda_i}{\partial \alpha} o_i(x)
\\ &= \sum_{i= 1}^n \frac{\partial\lambda_i}{\partial \alpha}\left[O_i - o_i(x)\right],
\end{split}
\enq
where the second line follows from the first after application of the chain rule and equation (\ref{lambda}).
Using the covariance matrix of the constraints given in (\ref{fluctuations}), the Fisher information matrix, defined in (\ref{FisherII}), can then be written as a double sum over the potentials,
\beq
\begin{split}\label{Zform}
F^X_{\alpha\beta} &= \sum_{i,j = 1}^n\frac{\partial\lambda_i}{\partial \alpha}\frac{\partial^2\ln Z}{\partial \lambda_i\partial\lambda_j} \frac{\partial\lambda_j}{\partial \beta}.
\end{split}
\enq
There are several ways to rewrite this expression as a function of the constraints and/or their potentials. First, it can be written as a single sum by using equation (\ref{lambda}) as
\beq \label{sumform}
F^X_{\alpha\beta} = -\sum_{i = 1}^n\frac{\partial\lambda_i}{\partial\alpha}\frac{\partial O_i}{\partial \beta}.
\enq
Alternatively, since we will be more interested in using the constraints as the main variables, and not the potentials,  we can show, using equation (\ref{potentials}), that it also takes the form \footnote{We note that this result is valid only for maximal entropy distributions and is not equivalent to the second derivative of the entropy with respect to the parameters themselves. However it is formally identical to the corresponding expression for the information content of distributions within the exponential family \citep{Jennrich75}, or \citep[chapter 4]{vandenbos07}, once the curvature of the entropy surface is identified with the generalized inverse of the covariance matrix.}
\beq \label{fishermaxent}
F^X_{\alpha\beta} = -\sum_{i,j = 1}^n\frac{\partial O_i}{\partial \alpha} \frac{\partial^2 S}{\partial O_i O_j}\frac{\partial O_j}{\partial \beta}.
\enq
We will use both of these last expressions in the following parts of this work.\newline
Equation (\ref{fishermaxent}) presents the total amount of information on the model parameters $\btheta$ in the data $X$, when the model predicts the set of constraints $O_i$. The amount of information is in the form of a sum of the information contained in each constraint, with correlations taken into account, as in the right hand side in equation (\ref{MultiCramer}). In particular, it is a property of the maximal entropy distributions, that if the constraints $O_i$ are not redundant, then it follows that the curvature matrix of the entropy surface $- \partial^2S$ is invertible and is the inverse of the covariance matrix $\partial^2 \ln Z $  between the observables. To see this explicitly,  consider the derivative of equation (\ref{lambda}) with respect to the potentials,
\beq
-\frac{\partial O_i }{\partial \lambda_j}  = \frac{\partial^2\ln Z}{\partial \lambda_i\partial\lambda_j}.
\enq
The inverse of the matrix on the left hand side, if it can be inverted, is $-\frac{\partial \lambda_i }{\partial O_j}$, which can be obtained taking the derivative of equation (\ref{potentials}), with the result
\beq
-\frac{\partial \lambda_i }{\partial O_j} =- \frac{\partial^2 S}{\partial O_i\partial O_j}.
\enq
We have thus obtained in equation (\ref{fishermaxent}), combining Jaynes' Maximal Entropy Principle together with Fisher's information, the exact expression of the Cram\'{e}r-Rao 
 inequality (\ref{MultiCramer}) for our full set of constraints,  but with an equality sign. \newline\newline
We see that the choice of maximal entropy probabilities is fair, in the sense that all the Fisher information comes from what was forced upon the probability density function, i.e. the constraints. No additional Fisher information is added when these probabilities are chosen. In fact, this requirement alone is enough to single out the maximal entropy distributions, as being precisely those for which the Cram\'{e}r-Rao 
 inequality is an equality. This can be understood in terms of sufficient statistics and goes back to \citep{1936PCPS...32..567P} and \cite{koopman36}. This was shown in \citep{RePEc:spr:metrik:v:41:y:1994:i:1:p:109-119}. We provide in appendix \ref{CramerRao bound} for completeness a similar argument that if the equality sign holds in equation (\ref{MultiCramer}) for some distribution, then it is the one that maximises the entropy relative to some other distribution. \newline\newline
In the special case that the model parameters are the constraints themselves, we have
\beq \label{entropysurface}
F^X_{O_iO_j}= - \frac{\partial^2 S}{\partial O_i O_j} = -\frac{\partial \lambda_i }{\partial O_j},
\enq
which means that the Fisher information on the model predictions contained in the expected future data is directly given by the sensitivity of their corresponding potential. Also, the application of the Cram\'{e}r-Rao 
 inequality, in the form given in equation (\ref{covF}), to any set of unbiased estimators of $\mathbf O$, shows that the best joint, unbiased, reconstruction of $\mathbf O$ is given by the inverse curvature of the entropy surface $-\partial^2S$, which is, as we have shown, $\partial^2 \ln Z$.\newline\newline
We emphasise at this point that although the amount of information is seen to be identical to the Fisher information in a Gaussian distribution of the observables with the above correlations, nowhere in our approach do we assume Gaussian properties. The distribution of the constraints $o_i(x)$ themselves is set by the  maximal entropy distribution of the data.\newline
\subsection{Redundant observables}\label{redundant}
We have just seen that in the case of independent constraints, the entropy of $p_X$ provides through equation (\ref{fishermaxent}) both the joint information content of the data, as well as the inverse correlation matrix between the observables. However, if the constraints put on the distribution are redundant, the correlation matrix is not invertible, and the curvature of the entropy surface cannot be inverted either. We show however that in these cases, our equations for the Fisher information content (\ref{Zform}, \ref{sumform}, \ref{fishermaxent}) are still fully consistent, dealing automatically with redundant information to provide the correct answer.\newline

\indent An example of redundant information occurs trivially if one of the functions $o_i(x)$ can be written in terms of the others. For instance, for galaxy survey data, the specification of the galaxy power spectrum as an constraint, together with the mean number of galaxy pairs as function of distance, and/or the two-points correlation function, which are three equivalent descriptions of the same statistical property of the data. Although the number of observables $\mathbf O$, and thus the number of potentials, describing the maximal entropy distribution greatly increases by doing so, it is clear that we should expect the Fisher matrix to be unchanged, by adding such superfluous pieces of information. A small calculation shows that the potentials adjust themselves so that it is actually the case, meaning that this type of redundant information is automatically discarded within this approach.  Therefore, we need not worry about the independency of the constraints when evaluating the information content of the data, which will prove convenient in some cases.\newline

\indent There is another, more relevant type of redundant information, that allow us to understand better the role of the potentials. Consider that we have some set of constraints $\{O_i\}_{i =1}^n$, and that we obtain the corresponding $p_X$ that maximises the entropy. This $p_X$ could then be used to predict the value $O_{n+1}$ of the average some other function $o_{n+1}(x)$, that is not contained in our set of predictions, 
\beq
\av{o_{n +1}(x)}  = : O_{n +1}.
\enq
For instance, the maximal entropy distribution built with constraints on the first $n$ moments of $p_X$, will predict some particular value for the $n+1$-th moment, $O_{n+1}$, that the model was unable to predict by itself.\newline
Suppose now some new theoretical work provides the shape of $O_{n+1}$ as a function of the model parameters. This new constraint can thus now be added to the previous set, and a new, updated $p_X$ is obtained by maximising the entropy. There are two possibilities at this point :
\newline
\newline \indent
It may occur that the value of $O_{n+1}$ as provided by the model is identical to the prediction by the maximal entropy distribution that was built without that constraint. Since the new constraint was automatically satisfied, the maximal entropy distribution satisfying the full set of $n +1$ constraints must be equal to the one satisfying the original set. From the equality of the two distributions, which are both of the form (\ref{MaxEnt}), it follows that the additional constraint must have vanishing associated potential,
\beq
\lambda_{n+1} = 0,
\enq
while the other potentials are pairwise identical. It follows immediately that the total information, as seen from equation (\ref{sumform}) is unaffected, and no information on the model parameters was gained by this additional prediction.
A cosmological example would be to enforce on the distribution of some field, together with the two-points correlation function, fully disconnected higher order correlation functions. It is well known that the maximal entropy distribution with constrained two-points correlation function has a Gaussian shape, and that Gaussian distributions have disconnected points function at any order. No information is thus provided by these field moments of higher order in this case.\newline
This argument shows that, for a given set of original constraints and associated maximal entropy distribution, any function $f(x)$, which was not contained in this set, with average $F$, can be seen as being set to zero potential. Such $F$'s therefore do not contribute to the information.
\newline\newline\indent
More interesting is, of course, the case where this additional constraint differs from the predictions obtained from the original set $\left\{O_i\right\}_{i = 1}^n$. Suppose that there is a mismatch $\delta O_{n+1}$ between the predictions of the maximal entropy distribution and the model. In this case, when updating $p_X$ to include this constraint, the potentials are changed by this new information, a change given to first order by
\beq
\delta\lambda_i = \frac{\partial^2S}{\partial O_i \partial O_{n+1}}\delta O_{n+1}, \quad  i = 1,\cdots, n+1,
\enq
and the amount of Fisher information changes accordingly.
\newline\newline
Of course, although the formulae of this section are valid for any model, it requires numerical work in order to get the partition function and/or the entropy surface in a general situation.
\subsection{The entropy and Fisher information content of Gaussian homogeneous fields} \label{Examples}
In order to close this section, we obtain now the Shannon entropy of a family of fields when only the two-point correlation function is the relevant constraint, that we will use extensively in the next section dealing with our cosmological application. It is easily obtained by a straightforward generalisation of the finite dimensional multivariate case, where the means and covariance matrix of the variables are known. It is well known  \citep{shannon48}  that the maximal entropy distribution is in this case the multivariate Gaussian distribution. Denoting the constraints on $p_X$ with the matrix $D$ and vector $\boldsymbol \mu$
\beq \label{cov} \begin{split}
D_{ij} &=  \av{x_{i}x_{j}} \\
\mu_i &= \av{x_i}, \quad i,j = 1,\cdots,N
\end{split}
\enq
the associated potentials are given explicitly by the relations
\beq \begin{split} \label{Dm}
 \lambda &= \frac 12 C^{-1}
\\\eta &= - C^{-1}\bmu,
\end{split}
\enq
where the matrix $C$ is the covariance matrix
\beq
C := D - \bmu \bmu^T.
\enq
The Shannon entropy is given by, up to some irrelevant additive constant,
\beq \label{entropyG}
S(D,\bmu) = \frac 12 \ln \det (D -  \bmu \bmu^T).
\enq
The fact that about half of the constraints are redundant, due to the symmetry of the $D$ and $C$ matrices, is reflected by the fact that the corresponding inverse correlation matrix in equation (\ref{fishermaxent}),
\beq
-\frac{\partial^2 S}{\partial D_{ij}\partial D_{kl}} = - \frac{\partial \lambda_{ij}}{\partial D_{kl}} =  \frac 12  C\inv_{ik}C\inv_{jl},
\enq
is not invertible as such if we considers all entries of the matrix $D$ as constraints. Of course,  this is not the case anymore if only the independent entries of $D$ form the constraints.  
\subsubsection{Fields, means and correlations}
\newcommand{\bbphi}{\bar\bphi}
Using  the handy formalism of functional calculus, we can straightforwardly extend the above relations to systems with infinite degrees of freedom, i.e. fields, where means as well as the two-point correlation functions are constrained. A realisation of the variable $X$ is now a field, or a family of fields $\bphi = (\phi_1,\cdots,\phi_N)$, taking values on some $n$-dimensional space.  The expressions above in the multivariate case all stays valid, with the understanding that operations such as matrix multiplications have to be taken with respect to the discrete indices as well as the continuous ones.\newline\newline
With the two-point correlation function and means
\beq
\begin{split}
\rho_{ij}(\vecx,\vecy) &= \av{\phi_i(\vecx)\phi_j(\vecy)} \\
\bar\phi_i(\vecx) &= \av{\phi_i(\vecx)}
\end{split}
\enq
we still have, up to an unimportant constant,
\beq \label{entropyfield}
S = \frac 12 \ln \det (\rho - \bphi\bphi^T). 
\enq
\newline\newline
In n-dimensional Euclidean space, within a box of volume $V$ for a family of homogeneous fields, it is simplest to work with the spectral matrices. These are defined as
\beq \label{spectralmatrix}
\frac 1 V \lat \tilde\phi_i(\veck) \tilde\phi_j^*(\veck')\rat =  P_{ij}(\veck)\:\delta_{\veck\veck'},
\enq
where the Fourier transforms of the fields are defined through
\beq
\tilde\phi_i(\veck) = \int_V d^nx \: \phi_i(x)\:e^{-i\veck\cdot\vecx}.
\enq
It is well known that these matrices provide an equivalent description of the correlations, since the they form Fourier pairs with the correlation functions
\beq
\rho_{ij}(\vecx,\vecy) = \frac 1 V \sum_{\veck} P_{ij}(\veck) e^{i\veck \cdot (\vecx - \vecy)} = \rho_{ij}(\vecx -\vecy).
\enq
In this case, the entropy in equation (\ref{entropyfield}) reduces, again discarding irrelevant constants, to an uncorrelated sum over the modes,
\beq \label{entropy field}
S = \frac 12 \ln\det\left[ \frac {P(0)}{V} - \bar\phi\bar\phi^T \right]+ \frac 12\sum_{\veck}\ln \det \frac{ P(\veck) }{V},
\enq
which is the straightforward mutlidimensional version of \citep[eq. 39]{2001MNRAS.328.1027T}.
Comparison with equation (\ref{entropyG}) shows the well-known fact that the modes can be seen as Gaussian, uncorrelated and complex variables with correlation matrices proportional to $P(\veck)$.  All modes have zero mean, except for the zero-mode, which, as seen from its definition, is proportional to the mean of the field itself. Accordingly, taking the appropriate derivatives, the potentials $\lambda(\veck)$ associated to $P(\veck) $ read
\beq
\begin{split} 
\lambda(\veck) &= \frac V{2} P(\veck) ^{-1},\quad \veck \ne 0 \\
\lambda(0) &=  \frac 1{2} \lb  \frac{P(0)}{V} - \bphi\bphi^T \rb^{-1} . 
\end{split}
\enq
and those associated to the means $\bphi$,
\beq
\boldsymbol \eta = - \lb  \frac{P(0)}{V} - \bphi\bphi^T \rb^{-1}\bphi
\enq
Note that although the spectral matrices are, in general, complex, they are hermitian, so that the determinants are real. The amount of Fisher information in the family of fields is easily obtained with the help of equation (\ref{sumform}) , with the familiar result
\beq \label{FPk} \begin{split}
\Fab &= \frac 1 2\sum_{\veck } \Tr\left[P_c^{-1}(\veck)\frac{\partial P_c(\veck)}{\partial \alpha}P_c^{-1}(\veck)\frac{\partial P_c(\veck)}{\partial \beta}\right]\\
&\quad+ \frac{\partial \bbphi^T}{\partial \alpha}\lb \frac{P_c(0)}{V}\rb^{-1} \frac{\partial\bbphi}{\partial \beta},
\end{split}
\enq
with $P_c(\veck)$ being the connected part of the spectral matrices,
\beq
\begin{split}
P_c(\veck) &= P(\veck) - \delta_{\veck0}V\bphi\bphi^T.
\end{split}
\enq
These expressions are of course also valid for isotropic fields on the sphere. With a decomposition in spherical harmonics, the sum runs over the multipoles.
\section{Cosmological application to weak lensing observables} \label{application}
Gravitational lensing, which can be used to measure the distribution of mass along the line of sight, has been recognized as powerful probe of the dark components of the Universe \citep{1992grle.book.....S,2001PhR...340..291B,2003ARA&A..41..645R,2006astro.ph.12667M,2006glsw.book.....S} since it is  sensitive to both the geometry of the Universe, and to the growth of structure. Weak lensing data is typically used in two ways. The first, which is deployed for cosmological parameter fitting, relies on measuring the correlated distortions in galaxy images \citep{2006astro.ph..9591A}. The second approach uses each galaxy to make a noisy measurement of the lensing signal at that position. These point estimates are then used to reconstruct the dark matter density distribution \citep[e.g.][]{1993ApJ...404..441K,2001A&A...374..740S}. Most of the measurements of weak lensing to date have focused on the shearing that galaxy images experience. However, gravitational lensing causes a number of other distortions of galaxy images. These include change in size, which is related to the magnification, and higher order image distortions known as the flexion \citep{2006MNRAS.365..414B}. A number of techniques have been developed for measuring these higher order images distortions, such as HOLICS \citep{2007ApJ...660..995O} and shapelets methods \citep{2007MNRAS.380..229M}. Since all of the image distortions originate from the same cause, i.e. the lensing potential field, the information content of any two lensing measurements must be degenerate. At the same time, since each method has different systematics and specific noise properties, combining multiple measurement can may bring substantial benefits. Some recent works have looked at the impact of combining shear and flexion measurements for mass reconstruction \citep{2010arXiv1008.3088E,2010ApJ...723.1507P,2010arXiv1011.3041V}, as well as the benefits for breaking multiplicative bias of including galaxy size measurements  \citep{2010arXiv1009.5590V}.
\black
\subsection{Linear probes}
The predictive power of some observable  $O_c$ of a central field (for instance its power spectrum at some mode)  translates into an array of constraints $O_i,\:\: i = 1,\cdots,n$ in the noisy probes, that we could try and extract and confront to theory :
\beq
\begin{split}\label{assum}
O_i(\btheta) = f_i(O_c(\btheta)), \quad i = 1,\cdots, n
\end{split}
\enq
for some functions $f_i$.\newline
For the purpose of this work,  the case of functions linear with respect to $O_c$ is generic enough, i.e. we will consider that
\beq
\frac{\partial^2 f_i}{\partial O_c^2} = 0,\quad i = 1,\cdots,n.
\enq
The entropy $S$ of the data is a function of the $n$ constraints $\mathbf O$. It is however fundamentally a function of $O_c$ since it does enter all of these observables. It is therefore very natural to associate a potential $\lambda_c$ to $O_c$, although it is not itself a constraint on the probability density function. In analogy with
\beq
\lambda_i =  \frac{\partial S}{\partial O_i}, \quad i = 1,\cdots,n
\enq
we define
\beq
\lambda_c := \frac{d S}{dO_c}(O_1,\cdots,O_m),
\enq
with the result, given by application of the chain rule, of
\beq  \label{un}
\lambda_c = \boldsymbol \lambda \cdot \frac{\partial \mathbf f}{\partial O_c}.
\enq
On the other hand, the impact of a model parameter on each observables can be similarly written in terms of the central observable $O_c$,
\beq \label{deux}
\frac{\partial \mathbf O}{\partial \alpha} = \frac{\partial O_c}{\partial \alpha} \frac{\partial \mathbf f}{\partial O_c} .
\enq
It follows directly from these relations (\ref{un}) and (\ref{deux}), and the linearity of $f_i$, that the joint information in the full set of constraints $\mathbf O$, given in equation (\ref{sumform}) as a sum over all $n$ constraints, reduces to a formally identical expression with the only difference that only $O_c$ enters :
\beq
\Fab^X = \frac{\partial \mathbf O}{\partial \alpha }\cdot \frac{\partial \boldsymbol \lambda}{ \partial \beta} = \frac{\partial \lambda_c}{\partial \alpha}\frac{\partial O_c}{\partial \beta},
\enq
which can also be written in the form analog to (\ref{fishermaxent}),
\beq  \label{infreduction}
\Fab^X =  -\frac{\partial O_c}{\partial \alpha}\frac{d^2 S}{dO_c^2} \frac{\partial O_c}{\partial \beta}.
\enq
This last equation shows that all the effect of combining this set of constraints have been absorbed into the second total derivative of the entropy. This second total derivative is the total amount of information there is on the central quantity $O_c$ in the data. Indeed, taking as a special case of model parameter to the central quantity itself, i.e. 
\beq
\alpha =  \beta =  O_c, 
\enq
one obtains now that the full amount of information in $X$ on $O_c$ is 
\beq \label{Infcentr}
F^X_{O_cO_c} = -\frac{d^2 S}{d O_c^2}\lp O_1,\cdots,O_n \rp \equiv \frac{1}{\sigma^2_{\eff}}.
\enq
A simple application of the Cramer Rao inequality presented in equation (\ref{CramerRao}) shows that this effective variance $\sigma_{\eff}$ is the lower bound to an unbiased reconstruction of the central observable from the noisy probes. \newline\newline
These considerations on the effect of probe combination in the case of a single central field observable $O_c$ generalize easily to the case where there are many, $(O_c^1,\cdots, O_c^m)$. In this case, each central field quantity leads to an array of constraints in the form of equation (\ref{assum}), it is simple to show that the amount of Fisher information can again be written in terms of the information associated to the central field, with an effective covariance matrix between the $O_c's$. The result is
\beq
\Fab^X = -\sum_{i,j = 1}^{m}\frac{\partial O_c^i}{\partial \alpha}\frac{d^2S}{dO_c^iO_c^j}\frac{\partial O_c^j}{\partial \beta}.
\enq
All the effects of probe combination are thus encompassed in  an effective covariance matrix $\Sigma_\eff$ of the central field observables,
\beq
-\frac{d^2S}{dO_c^iO_c^j} \equiv \lb \Sigma^{-1}_\eff\rb_{ij}.
\enq
Again, an application of the Cramer Rao inequality, in the multi-dimensional case, shows that this effective covariance matrix is the best achievable unbiased joint reconstruction of  $(O_c^1,\cdots, O_c^m)$.
\newline\newline
We now explore further the case of linear probes of homogeneous Gaussian fields, which is cosmologically relevant and can be solved analytically to full extent. We will focus on zero mean fields,  for which according to our previous section the entropy can be written in terms of the spectral matrices, up to a constant,
\beq \label{EntropyField}
S = \frac 12 \sum_{\veck} \ln \det P(\veck).
\enq
\color{black}
\subsection{Linear tracers at the two-point level}
A standard instance of a linear tracer $\phi_i$ of some central field $\kappa$ in weak lensing is provided by a relation in Fourier space of the form
\beq \label{example}
\tilde \phi_i(\veck) = v_i \tilde \kappa(\veck) +\tilde \epsilon_i(\veck),
\enq
for some noise term $\tilde\epsilon_i$, uncorrelated with $\kappa$, and coefficient $v_i$. Typically, if one observes a tracer of the derivative of the field $\kappa$, then the vector $\vecv$ would be proportional to $-i\veck$. We are ignoring here any observational effect, such as incomplete sky coverage, that would require corrections to this relation.  It is clear from this relation that the spectral matrices of this family take a special form of equation (\ref{assum}): defining the spectrum of the $\kappa$ field by $P^\kappa$, we obtain by putting this relation (\ref{example}) into (\ref{spectralmatrix}), that the spectral matrices can be written at each mode in the form
\beq \label{specmatrix}
P =  P^\kappa \vecv\vecv^\dagger + N,
\enq
where $\vecv^\dagger$ is the hermitian conjugate of $\vecv = \lp v_1,\cdots,v_n\rp$. The matrix $N$ is the spectrum of the noise components $\epsilon$,
\beq
N_{ij}(\veck) = \frac 1 V\av{\tilde \epsilon_i(\veck)\tilde \epsilon^*(\veck)}.
\enq 
Our subsequent results hold for any family of tracers that obey this relation. While the special case in (\ref{example}) enter this category, this need not be the only instance. All the weak lensing observables we deal with in this work will satisfy equation (\ref{specmatrix}).
\newline\newline
Both the n-dimensional vector $\vecv$ and the noise matrix $N$ can depend on the wave vector $\veck$, but they are independent from the model parameters. The  matrix $N$ of dimension $n\times n$ is the noise component of the spectra of the fields, typically built from two parts. The first is due to the discrete nature of the fields, since such data consist in quantities measured where galaxies sits, and the second to the intrinsic dispersion of the measured values.
\subsection{Joint entropy and information content}
\indent Information on the model parameters enters through $P^\kappa$ only.
To evaluate the full information content, we need only evaluate eq. (\ref{EntropyField}) with the spectral matrix given in (\ref{specmatrix}), keeping in mind the result from last section, that we need only the total derivative with respect to $P^\kappa$. In other words, any additive terms in the expression of the entropy that are independent of $P^\kappa$ can be discarded.
\newline\newline
This determinant can be evaluated immediately. Defining for each mode the real positive number $N_\eff$ through
\beq \label{noise}
\frac 1 {N_\eff} \equiv \vecv^\dagger N^{-1}\vecv,
\enq
which can be seen as an effective noise term, a simple\footnote{We have namely for any invertible matrix $A$ and vectors $\vecu,\vecv$ the matrix determinant lemma,
\beq
\det \left(A + \vecu \vecv^T\right) = \det\lp A\rp \lp 1 + \vecv^TA^{-1}\vecu \rp. 
\enq}  calculation shows that the joint entropy (\ref{EntropyField}) is equivalent to the following, where the $n$ dimensional determinant has disappeared,
\beq \label{Entropy2}
S = \frac12\sum_{\veck}\ln \left(P^\kappa(\veck)  +  N_\eff(\veck)\right).
\enq
Comparison with equation (\ref{EntropyField}) shows that we have with this equation (\ref{Entropy2}) the entropy of the field $\kappa$ itself, where all the effects of the joint observation of this $n$ fields have been absorbed into the effective noise term $N_\eff$, that contaminates its spectrum. It means that the full combined information in the $n$ probes of the field $\kappa$ is equivalent to the information in $\kappa$, observed with spectral noise $N_\eff$.\newline\newline
Our result (\ref{Infcentr}) applied to (\ref{Entropy2}) puts  bounds on reconstruction of the field $\kappa$ out of the observed samples, which can be at best reconstructed with a contaminating noise term of $N_\eff$ in its spectrum, whose best unbiased reconstruction is given by
\beq
 2\lp P^\kappa(\veck) +N_\eff(\veck)\rp^2.
\enq
\newline
Since the effect of combining these probes at a single mode is only to change the model independent noise term, the parameter correlations and degenaracies as approximated by the Fisher information matrix stay unchanged, whatever the number of such probes is. We have namely from (\ref{Entropy2}) that at a given mode $\veck$, the Fisher information matrix reads
\beq \label{fmatrixpk}
\Fab^X = \frac 12 \frac{\partial \ln \tilde P^\kappa(\veck) }{\partial\alpha}\frac{\partial \ln \tilde P^\kappa(\veck)}{\partial\beta},
\enq
with
\beq
 \tilde P^\kappa(\veck) =  P^\kappa(\veck)  + N_\eff(\veck).
\enq
From the point of view of the Fisher information, it makes formally no difference to extract the full set of $n(n-1)/2$ independent elements of each spectral matrices, or reconstruct the field $\kappa$ and extract its spectrum. They carry indeed the same amount of Fisher information.
\newline\newline
These results still hold when other fields are present in the analysis, which are correlated 
with the field $\kappa$. To make this statement rigorous, consider in the analysis on top of our $n$ samples of the form (\ref{example}) of $\kappa$, another homogeneous field $\theta$, with spectrum $P^\theta(\veck)$, and cross spectrum to $\kappa$ given by $P^{\theta\kappa}(\veck)$ The full spectral matrices are in this case
\beq
P(\veck) = \bem P^\kappa(\veck) \vecv\vecv^T + N  & P(\veck)^{\kappa\theta}\vecv 
\\ P^{\theta\kappa}\vecv^T & P^{\theta}(\veck) \enm.
\enq
Again, the determinant of this matrix can be reduced to a determinant of lower 
dimension, leading to the equivalent entropy
\beq \label{multis}
S = \cst + \frac 12 \ln \det \bem P^\psi(\veck) + N_\eff & P^{\kappa\theta}
(\veck) \\ P^{\theta\kappa}(\veck) & P^\theta(\veck) \enm.
\enq
It shows that the the full set of $n + 1$ 
fields can be reduced without loss to two fields, $\kappa$ and 
$\theta$, with the effective noise $N_\eff$ contaminating the spectrum of $\kappa$. \newline\newline
Note that the derivation of our results do not refer to any hypothetical estimators, but 
came naturally out of the expression of the entropy.
\subsection{Weak lensing probes}
We now seek a quantitative evaluation of the full joint information content of the weak lensing probes in galaxy surveys, up to second order in the image distortions of galaxies. The data $X$ consists of a set of fields, which are discrete point fields, which take values where galaxies sits. We work in the two-dimensional flat sky limit, using the more standard notation $\mathbf l$ for the wave vector, and decompose it in modulus and polar coordinate as
\beq
\vecl = l\bem \cos \varphi_l \\ \sin \varphi_l \enm
\enq
For the scope of this paper, we will throughout assume that the intrinsic values of each probe are pairwise uncorrelated, as commonly done. Also, we will assume that the set of points on which the relevant quantities are measured show low enough clustering so that corrections to the spectra due to intrinsic clustering can be neglected. This is however not a limitation of our approach, since corrections to the above assumptions, such as the introduction of some level of intrinsic alignment, can be accommodated for by introducing appropriate terms in the noise matrices $N(\veck)$ in (\ref{noise}). 
As a central field to which all our point fields relates, we take for convenience the isotropic convergence field $\kappa$, with spectrum
\beq
C^\kappa(\vecl) = C^\kappa(l).
\enq
In the case of pairwise uncorrelated intrinsic values that we are following, we see easily from (\ref{noise}) that by combining any number of such probes the effective noise is reduced at a given mode according to
\beq \label{combination}
\frac 1 {N^{\textrm{tot}}_\eff} = \sum_i \frac{1}{N_\eff^i}.
\enq
We therefore only need to evaluate the effective noise for each probe separately, while their combination follows (\ref{combination}). To this aim, the evaluation of the spectral matrices (\ref{specmatrix}), giving us $N_\eff$, is necessary. The calculations for this are presented in appendix \ref{pf} and we use the final results in this section.
\subsubsection{First order, distortion matrix}
To first order, the distortion induced by weak lensing on a galaxy image is described by the distortion matrix that contains the shear, $\gamma$, and convergence, $\kappa$, which come from the second derivatives of the lensing potential field $\psi$, \citep[e.g. ][]{2006glsw.book.....S}
\beq \label{distortion}
\bem \kappa + \gamma_1 & \gamma_2 \\ \gamma_2 & \kappa - \gamma_1 \enm = \psi_{,ij}.
\enq
The shear components read
\beq
\gamma_1 = \frac 12 \left(\psi_{,11} - \psi_{,22}\right), \quad \gamma_2 = \psi_{,12}
\enq
and we assume they are measured from the apparent ellipticities of the galaxies, with identical intrinsic dispersion $\sigma^2_\gamma$. Denoting with $\bar n_\gamma$ the number density of galaxies for which ellipticity measurements are available,
the effective noise is 
\beq \label{neffgamma}
N^\gamma_\eff = \frac{\sigma^2_\gamma}{\bar n_\gamma}.
\enq
The information content of the two observed ellipticity fields is thus exactly the same as the one of the convergence field, with a mode independent noise term as above. 
\newline\newline
\newcommand{\intr}{\textrm{int}}
To reach for the $\kappa$ component of the distortion matrix, we imagine we have measurements of their angular size $s_\obs$, with intrinsic dispersion $\sigma^2_s$. The intrinsic sizes of the galaxies $s_\intr$ gets transformed through weak lensing according to
\beq
s_\obs = s_{\textrm{int}}( 1 +\alpha_s\kappa).
\enq 
The coefficient $\alpha_s$, is equal to unity in pure weak lensing theory, but we allow it to take other values, since in a realistic situation, other effects such as magnification bias effectively enter this coefficient (e.g. \citep{2010arXiv1009.5590V}). Under our assumption that the correlation of the fluctuations in intrinsic sizes can themselves be neglected, the effective noise reduces to
\beq\label{neffsize}
 N^s_\eff =  \frac 1 {\alpha_s^2}\lp \frac{\sigma_s} { \bar s_\textrm{int}} \rp ^2\frac{1}{\bar n_s}.
\enq
This combination of $\alpha_s$ with the dispersion parameters $\bar s$ and $\sigma_s$ becomes the only relevant parameter in our case, and not the value of each of them.
\subsubsection{Second order, flexion}
To second order, the distortions caused by lensing on the galaxies images are given by third order derivatives of the lensing potential. These are conveniently described by the spin 1 and spin 3 flexion components $\mathcal F$ and $\mathcal G$, which in the notation of  \citep{2008A&A...485..363S} read
\beq \begin{split}
\mathcal F &= \frac 12 \bem \psi_{,111} + \psi_{,122} \\ \psi_{,112} + \psi_{,222} \enm 
\\
\mathcal G &= \frac 12 \bem \psi_{,111} - 3\psi_{,122} \\ 3\psi_{,112} - \psi_{,222} \enm,
\end{split}
\enq
and are extracted from measurements with intrinsic dispersion $\sigma^2_\mathcal F$ and $\sigma^2_\mathcal G$.
The effective noise is this time mode-dependent,
\beq \label{neffflex}
\frac 1 { N^{\mathcal F \mathcal G}_\eff} = l^2\left(\frac{\bar n_F}{\sigma_F^2} + \frac{\bar n_G}{\sigma_G^2}\right).
\enq
\section{Results} \label{results}
\begin{center}
\begin{figure} 
  \includegraphics[width = 0.45\textwidth]{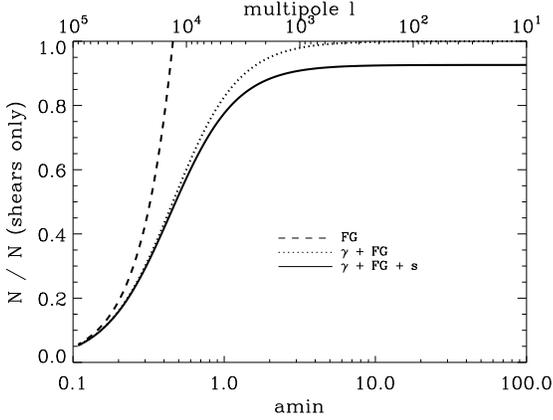}
  \caption{The ratio of the effective noise to the level of noise considering the shears only, as function of angular scale. The dashed line considers the flexion fields alone. The dotted line shows the combination of the flexion fields with the shear fields,  and the solid line all these weak lensing probes combined. No correlations between the intrinsic values for each pair of probes have been considered.}
  \label{noiseratio}
\end{figure}
\end{center}

\begin{center}
\begin{figure} 
  \includegraphics[width = 0.47\textwidth]{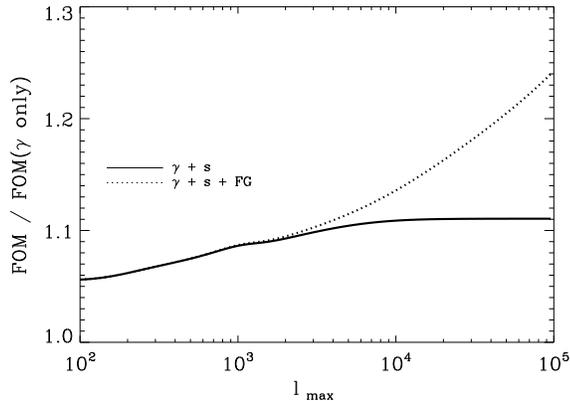}
  \caption{The improvement of the dark energy FOM including size information (solid), as well as flexion $\mathcal F$ and $\mathcal G$ information (dotted), over the shear only analysis, as function of the maximal angular multipole included in the analysis.}
  \label{figFOM}
\end{figure}
\end{center}

Figure \ref{noiseratio} shows the ratio of the effective noise to the noise present considering the shear fields only, assuming the same number densities of galaxies for each probe, and the values for the intrinsic dispersion stated in table \ref{paramtable}. The conversion multipole $l$ (upper x-axis) to angular scale $\theta$ (lower x-axis) follows $\theta = \pi / (l + 1/2)$.  We have adopted for the size dispersion parameters the numbers from \citep{2010arXiv1009.5590V}, who evaluated this number for the DES survey conditions \citep{2005astro.ph.10346T}. 
We refer to the discussion in \citep{2010ApJ...723.1507P} for our choice of flexion dispersion parameters.
The curves on this figure are rations and therefore independent of the galaxy number density. They are redshift independent as well, only to the extent that the dispersion in intrinsic values can be treated as such. We can draw two main conclusions from figure \ref{noiseratio}. First, flexion information beings to play role only at the smallest scales, on the arcsecond scales, where it takes over and becomes the most interesting probe. On the scale of $1$ amin, it can bring substantial improvement over shear only analysis, but only in combination with the shears, and not on its own. This is in good agreement with the comparative analysis of the power of the flexion $\mathcal F$ field and shears fields for mass reconstruction done in \cite{2010ApJ...723.1507P}, restricted to direct inversion methods. Second, the inclusion of size of galaxies into the analysis provides a density independent, scale independent, improvement factor of
\beq \label{improve}
\frac{N_\eff^\gamma}{N_\eff^{\gamma+ s}} = 1 + \lp \frac{\sigma_\gamma \bar s}{\sigma_s \alpha_s}\rp^2,
\enq
which is close to a $10\%$ improvement for the quoted numbers. Of course, the precise value depends on the dispersion parameters of the population considered.
\begin{center}
\begin{table} \caption{Dispersion parameters used in figure \ref{noiseratio}.}\label{paramtable}
\begin{tabular*}{0.3\textwidth}{cccc} 
 $\sigma_\gamma$ & $ \sigma_\mathcal F$ asec$^{-1}$ &$ \sigma_\mathcal G $ asec$^{-1}$ &  $\frac 1 {\alpha_s} \frac{ \sigma_s}{ \bar s}$ \\ \hline
 0.25 & 0.04 & 0.04 & 0.9 
 \end{tabular*}
\end{table}
\end{center}
For the purpose of measuring cosmological parameters rather than mass reconstruction, more interesting are the actual values of the Fisher information matrices. Since with any combination of such probes, these matrices are proportional to each other at a single mode, it makes sense to define the efficiency parameter of the probe $i$ through
\beq
\epsilon_i(l) := \frac{C^\kappa(l)}{C^\kappa(l) + N^i_\eff(l)},
\enq
which is a measure of what fraction of the information contained in the convergence field is effectively catched by that probe. The information in the convergence field is, at a given mode $l$, counting the multiplicity of the mode,
\beq
F^\kappa_{\alpha\beta} = \frac 12 (2l + 1) \frac{\partial \ln C^\kappa(l)}{\partial\alpha}\frac{\partial \ln C^\kappa(l)}{\partial\beta},
\enq
and we have indeed that the total Fisher information in the observed fields is
\beq
\Fab^X = \sum_l  F^\kappa_{\alpha\beta}(l) \epsilon^2_i(l).
\enq
Therefore, according to the interpretation of the Fisher matrix approximating the expected constraints on the model parameters, the factor $\epsilon(l)$ is precisely equal to the factor of degradation in the constraints one would be able to put on any a parameter, with respect to the case of perfect knowledge of the convergence field at this mode.
It is not the purpose of this work to perform a very detailed study on the behavior of the efficiency parameter for some specific survey and the subsequent statistical gain, but its qualitative behavior is easy to see. This parameter is essentially unity in the high signal to noise regime, while it is the inverse effective noise whenever the intrinsic dispersion dominates the observed spectrum. Since information on cosmological parameters is beaten down by cosmic variance in the former case, the latter dominates the constraints. We can therefore expect from our above discussion the size information to tighten by a few percent constraints on any  cosmological parameter. On the other hand, while flexion becomes ideal for mass reconstruction purposes on small scales, it will be able to help inference on cosmological parameters only if the challenge of  very accurate theoretical predictions on the convergence power spectrum for multipoles substantially larger than $1000$ will be met.\newline\newline
To make these expectations more concrete, we evaluated the improvement in information on cosmological parameters performing a lensing Fisher matrix calculation for a wide, EUCLID-like survey, in a tomographic setting. For a data vector consisting of $n$ probes of the convergence field $\kappa_i$ in each redshift bin $i,\: i = 1,\cdots N$, it is simple to see following our previous argument, that the Fisher information reduces to
\beq
F_{\alpha\beta} = \frac 12  \sum_{l}\lp 2l + 1\rp \Tr \:C^{-1}\frac{\partial C}{\partial\alpha} C^{-1}\frac{\partial C}{\partial\beta},
\enq
where the $C$ matrix is given by
\beq \label{cmatrix}
C_{ij} = C^{\kappa_i \kappa_j}(l)  + \delta_{ij}N^i_\eff(l),\quad i,j = 1,N
\enq
with $N_\eff^i$ given by (\ref{noise}). The only difference between standard implementations of Fisher matrices for lensing, such as the lensing part of \cite{2004PhRvD..70d3009H}, being thus the form of the noise component.
we evaluated these matrices respectively for
\beq \label{shearonly}
N_\eff^i = \frac{\sigma^2_\gamma}{\bar n^i} = N^{\gamma,i}_\eff ,
\enq
which is the precise form of the Fisher matrix for shear analysis, for
\beq\label{shearsize}
\frac{1}{N_\eff^i} =\frac 1 {N_\eff^{\gamma,i}} + \frac 1 {N_\eff^{s,i}}
\enq
which account for size information, and
\beq\label{shearsizefg}
\frac{1}{N^i_\eff(l)} = \frac 1 {N_\eff^{\gamma,i}} + \frac 1 {N_\eff^{s,i}} + \frac 1 {N_\eff^{\mathcal F \mathcal G,i}(l)} ,
\enq
which accounts for the flexion fields as well.  We note that in terms of observables, these small modifications incorporate in its entirety the full set of all possible correlations between the fields considered. The values of the dispersion parameters involved in these formulae are the same as in table (\ref{paramtable}). Our fiducial model is a flat $\Lambda$CDM universe, with parameters $\Omega_\Lambda = 0.7$, $\Omega_b = 0.045$, $\Omega_m = 0.3, h = 0.7$, power spectrum parameters $\sigma_8 = 0.8, n = 1$, and Linder's parametrisation \citep{2003PhRvL..90i1301L} of the dark energy equation of state implemented as $\omega_0 = -1,w_a = 0$. The distribution of galaxies as function of redshift needed both for the calculation of the spectra and to obtain the galaxy densities in each bin was generated using the cosmological package iCosmo \citep{2008arXiv0810.1285R}, in a way described in \citep{2007MNRAS.381.1018A}. We adopted EUCLID-like parameters of $10$ redshift bins, a median redshift of $1$, a galaxy angular density of $40 / \textrm{amin}^2$, and photometric redshift errors of $0.03(1 + z)$.
\newline\newline
In figure \ref{figFOM}, we show the improvement in the dark energy Figure of Merit (FOM), defined as the square root of the determinant of the submatrix $\lp \omega_0,\omega_a\rp$ of the Fisher matrix inverse $F^{-1}_{\alpha\beta}$ ($\alpha$ and $\beta$ running over the set of eight parameters as described above), as function of the maximal angular mode $l_\max$ considered, while $l_\min$ being always taken to be 10. In perfect agreement with our discussion above, including size information (solid line) increases the FOM steadily until it saturates at a $10\%$ improvement when constraints on the dark energy parameters are dominated by the low signal to noise regime. Also, flexion becomes only useful in the deep non-linear regime, where however theoretical understanding of the shape of the spectra still leaves a lot to be desired.
\newline\newline
These results are found to be very insensitive to the survey parameters, for a fixed $\alpha_s$.  There are also only weakly model parameter independent, as illustrated in table \ref{ratiotable}, which shows the corresponding improvement in Fisher constraints,
\beq
\frac{\sigma^2}{\sigma^2_{\textrm{shear only}}} = \frac{F^{-1}_{\alpha\alpha}}{F^{-1}_{\alpha\alpha, \textrm{shear only}}},
\enq
at the saturation scale $l_\max = 10^4$.
These results are also essentially unchanged using either standard implementations of the halo model \citep[for a review]{2002PhR...372....1C} or the the HALOFIT \citep{2003MNRAS.341.1311S} non linear power spectrum.
\begin{center}
\begin{table} \caption{Ratio of the marginalised constraints $\sigma^2 / \sigma^2_{\textrm{shear only}} $ , for $l_\max = 10^4$. This first line considers the inclusion of the size information in the analysis, while the second the size as well as the flexion fields $\mathcal F$ and $\mathcal G$.}.\label{ratiotable}
\begin{tabular*}{0.45\textwidth}{|cccccccc|}
 $\Omega_{\Lambda}$ & $\Omega_b$ & $\Omega_m$  & $h$  & $n$ &  $\sigma_8$ & $w_0$ & $w_a$ \\  \hline
 0.90 & 0.96 & 0.90 & 0.95 & 0.95 & 0.90 & 0.90 & 0.90 \\
  0.88 & 0.96 & 0.89 & 0.95 & 0.93 & 0.88 & 0.88 & 0.88
 \end{tabular*}
\end{table}
\end{center}
\black
\section{Summary and discussion}\label{discussion}
We have shown how Jaynes'  Maximal Entropy Principle allows us to construct the Fisher information content on model parameters in a given data set in the form of equation (\ref{sumform}) or (\ref{fishermaxent}). This is done by making the key quantity the entropy of the distribution as function of the constraints that we put on it. These constraints form our knowledge of the statistical properties of the future data. To the best of the authors knowledge,  equation (\ref{sumform}) or (\ref{fishermaxent}) are not to be found in this form in the literature. However, they cannot  be considered new, since as stated earlier, they can be easily gained from the Fisher information content of the exponential family of distributions \citep{Jennrich75,vandenbos07}, after the identification of the curvature of the entropy surface with the generalised inverse of the covariance matrix. Especially, the maximal entropy distributions are precisely those for which the Cram\'{e}r-Rao 
 inequality is an equality, since the curvature of the entropy surface is the inverse correlation matrix between the model predictions. Equation (\ref{fishermaxent}) also bears a strong formal similarity to the well known result  \cite[chap. 2]{Kullback} or \cite[chap. 6]{caticha08}, that the Fisher information can always be written as the curvature of the Kullback-Leibler divergence for distributions parametrised by the same set of parameters.
\newline\newline
The Fisher matrices currently used in weak lensing or clustering can all be seen as special cases of this approach, namely equation (\ref{FPk}), when knowledge of the statistical properties of the future data does not go beyond the two-point statistics. Indeed, in the case that the model does not predict the means, and knowing that for discrete fields the spectral matrices, equation (\ref{spectralmatrix}), carry a noise term due to the finite number of galaxies, or, in the case of weak lensing, also due to the intrinsic ellipticities of galaxies, the amount of information in (\ref{FPk})  is essentially identical to the standard expressions used to predict the accuracy with which parameters will be extracted from power-spectra analysis.
\newline\newline
There is, however, a conceptual difference worth noting in that the standard approach is to pick an estimator for the power-spectra and assume that both the fields as well as the distribution of the estimators are Gaussian. The result is the amount of Fisher information there is in the power spectra, under the assumptions of Gaussian statistics for the estimators and the fields. In our approach, the only assumption is on the fields distribution. Our results do not depend on the way the information will be extracted, but shows the amount of Fisher information in the fields as a whole. 
\newline\newline
Of course, the maximal entropy approach, which tries to capture the relevant properties of $p_X$ through a sophisticated guess, gives no guaranties that its predictions are actually correct. Nevertheless, as discussed in section \ref{redundant}, it provides a systematic approach with which to update the probability density function in case of improved knowledge of the relevant physics.
\newline
\newline
Using this formalism we have investigated the combined Fisher information content of weak lensing probes up to second order in the shapes distortions, assuming model parameter independent noise. By having a look at the joint Shannon entropy of the fields, we have shown how the only effect of treating these observables jointly is to reduce the effective level of noise contaminating the convergence field, according to equations (\ref{noise}) and (\ref{Entropy2}), independently of the model parameters.
\newline\newline\indent
These are the key points of this paper that we would like to emphasize :
\begin{enumerate}
\item Equation (\ref{fishermaxent}) presents a measure of information content that depends only on the constraints put on the data and the physical model. It is written in terms of the curvature of Shannon's entropy surface for maximal entropy distributions. It can always be interpreted, regardless of the actual distribution of the parameters, and of the specific way the analysis will proceed, as the expected curvature of the $\chi^2$ surface to the full set of model predictions. Assumptions of Gaussianity are neither needed nor used at any point. 
\item  Over a very wide range of scales, the probe of choice both for mass reconstruction or cosmological purposes are the ellipticity components of the galaxies. Flexion takes over only on the arcsecond scale. In combination with the ellipticities, it can lead to substantial increase in statistical power on the scale of the arcminute. From the cosmological point of view, we expect size information to contribute at the $10\%$ level of the total information content. The only key parameter is the combination (\ref{improve}) of the dispersion values and the permeability $\alpha_s$ of the population sizes to the convergence field. On the other hand, the prospects of including flexion in cosmological analysis are less clear. The most obvious drawback being the need for an accurate understanding of the non-linear power spectrum.
\item
Besides, our results render the inclusion of flexion and size information within more detailed Fisher matrix analysis for future dark energy experiments extremely simple, such as in the exhaustive approach combining the information galaxy density fields with shear fields in a tomographic setting of \citep{2009ApJ...695..652B}. From (\ref{multis}) follows namely that the inclusion of all the two-point correlations of these additional weak lensing probes can be accounted for by adapting the noise term $N_\eff$.
\end{enumerate}
\black
Possible developments of this work includes the relaxation of the main limitation of the results, for instance the assumption that the noise is independent of the model parameters. Also, we plan to show that the approach presented in the first part of this work can lead to quantitative evaluations in non Gaussian cases as well, when other observables than the first two moments are considered. 
\section*{Acknowledgments}
We thank Baptiste Schubnel for interesting discussions and a careful reading of the manuscript. We wish to thank also the anonymous referee for his detailed and useful comments on the manuscript. JC acknowledge the support of the Swiss National Science Foundation. AA is supported by
the Zwicky Fellowship at ETH Zurich. 
\begin{appendix}
\section{Cram\'{e}r-Rao 
 inequality}\label{CramerRao bound}
In this section, we provide a unified derivation of the Cram\'{e}r-Rao inequality in the multidimensional case (based on \citep{Rao}) and its relation to maximal entropy distributions. We denote the vector of model parameters of dimension $n$ with
\beq
\vecalpha = \bem\alpha_1,\cdots,\alpha_n\enm
\enq
and a vector of functions of dimension $m$ the estimators
\beq
\hat {\mathbf D} = \lp \hat D_1,\cdots \hat D_m \rp,
\enq
with expectation values $D_i(\vecalpha)=  \av {\hat D_i(x)} $. 
In the following, we rely on Gram matrices, whose elements are defined by scalar products. Namely, for a set of vectors $\vecy_i$, the Gram matrix $Y$ generated by this set of vectors is defined as
\beq
Y_{ij} = \vecy_i \cdot \vecy_j.
\enq
Gram matrices are positive definite and have the same rank as the set of vectors that generate them. Especially, if the vectors are linearly independent, the Gram matrix is strictly positive definite and invertible. \newline
We adopt a vectorial notation for functions, writing scalar products between vectors as
\beq
f\cdot g \equiv \int dx\:p_X(x,\vecalpha) f(x) g(x),
\enq
with $p_X(x,\vecalpha)$ being the probability density function of the variable $X$ of interest.
In this notation, both the Fisher information matrix and covariance matrix are seen to be Gram matrices. We have namely that the Fisher information matrix reads
\beq
 F_{ {\alpha_i}{\alpha_j}} = f_{\alpha_i} \cdot f_{\alpha_j},\quad f_{\alpha_i}(x) = \frac{\partial \ln p_X(x,\vecalpha)}{\partial \alpha_i},
\enq
while the covariance matrix of the estimators is
\beq
C_{ij} = g_i\cdot g_j,\quad g_i(x,\vecalpha) = \hat D_i(x) - D_i(\vecalpha). 
\enq
For simplicity and since it is sufficently generic for our purpose, we will assume that both sets of vectors $f$ and $g$ are lineary independent, so that both matrices can be inverted. Note that we also have
\beq
\frac{\partial D_i}{\partial {\alpha_j}} = \int dx \:p_X(x,\vecalpha) \:\hat D_i(x) \frac{\partial \ln p_X(x,\vecalpha)}{\partial {\alpha_j}} = g_i\cdot f_{\alpha_j}.
\enq
\newline
The  Gram matrix $ G$ of dimension $\lp (m + n) \times (m + n) \rp$ generated by the set of vectors $\lp g_1,\cdots, g_m ,f_{\alpha_1},\cdots, f_{\alpha_n}\rp $ takes the form
\beq
G = \bem  C & \Delta \\  \Delta^T & F \enm , \quad \Delta_{i {\alpha_j}} = g_i \cdot f_{\alpha_j}
\enq
and is also positive definite due to its very definition. It is congruent to the matrix
\beq
 Y G Y^T = \bem  C - \Delta F \inv \Delta^T & 0 \\ 0 & F \enm,
\enq
with
\beq
Y = \bem 1_{m\times m} & - \Delta F^{-1} \\ 0 & 1_{n\times n} \enm.
\enq
Since two congruent matrices have the same number of positive, zero and negative eigenvalues respectively and since both $ F$ and $G$  are positive, we can conclude that
\beq
C \ge \Delta F \inv \Delta^T,
\enq 
which is the Cram\'{e}r-Rao 
 inequality. The lower bound on the amount of information is seen from the fact that for any matrix written in block form holds
\beq
\bem  C & \Delta \\  \Delta^T & F \enm\ge 0 \Leftrightarrow \bem F & \Delta^T \\  \Delta & C \enm \ge 0
\enq
and using the same congruence argument leads to the lower bound on information
\beq
F \ge \Delta^T C \inv \Delta.
\enq
Assume now that we have a probability density function such that this inequality
is in fact an equality, i.e.
\beq
F = \Delta^T C \inv \Delta.
\enq
By the above argument, the Gram matrix generated by
\beq
\lp f_{\alpha_1},\cdots, f_{\alpha_n}, g_1,\cdots, g_m\rp
\enq
is congruent to the matrix
\beq
\bem 0_{n \times n} & 0 \\ 0& C \enm
\enq
and has rank $m$. By assumption, the covariance matrix is invertible, such that the set $\lp g_1,\cdots, g_m \rp$ alone has rank $m$. It implies that each of the $f$ vector can be written as linear combination of the $g$ vectors,
\beq
f_{\alpha_i} = \sum_{j = 1}^{m} A_{j} g_j,
\enq
or, more explicitly,
\beq
\frac{\partial \ln p_X(x,\vecalpha)}{\partial {\alpha_i}} =  \sum_{j = 1}^{m} A_{ j}(\vecalpha) \left[\hat{ D_j}(x) - D_j(\vecalpha)\right],
\enq
where the key point is that the coefficients $A_j$ are independent of $x$. Integrating this equation, we obtain
\beq
\ln p_X(x,\vecalpha) = -\sum_{i = 1}^m \lambda_i(\vecalpha)\hat {D_i}(x) - \ln Z(\vecalpha) + \ln q_X(x)
\enq
for some functions $\lambda$ and $Z$ of the model parameters only, and a function $q_X$ of $x$ only. We obtain thus
\beq
p_X(x,\vecalpha) = \frac {q_X(x)} {Z(\vecalpha)} \exp\left(-\sum^m_{i=1} \lambda_i (\vecalpha)\hat {D}_i(x) \right).
\enq
This is precisely the distribution that we obtain by maximising the entropy relative to $q_X(x)$, while satisfying the constraints
\beq
D_i(\vecalpha) = \av{\hat D_i(x)},\quad i = 1,\cdots, m.
\enq 
Taking $q_X$ as the uniform distribution makes it identical with the formula in equation (\ref{MaxEnt}).
\section{Point fields}\label{pf}
The data consists in a set of numbers, at each position where a galaxy sit and a measurement was done. We use the handy notation in terms of Dirac delta function,
\beq
\phi(\vecx) = \sum_i \epsilon_i\delta^D(\vecx - \vecx_i),
\enq
where the sum runs over the positions $\vecx_i$ for which $\epsilon$ is measured. To obtain the spectral matrices, we need the Fourier transform of the field, which reads in our case
\beq
\tilde \phi(\vecl) = \sum_i \epsilon_i \exp\lp - i \vecl\cdot \vecx_i\rp. 
\enq
In this work, we assume that the set of points shows negligible clustering, so that the probability density function for the joint occurrence of a particular set of galaxy positions is uniform.\newline
We decompose in the following the wave vector $\veck$ on the flat sky in terms of its modulus and polar angle as
\beq
\vecl = l\bem \cos\phi_l \\ \sin \phi_l \enm.
\enq
\subsection{Ellipticities}
When the two ellipticity components are measured, we have two such fields $\phi_1,\phi_2$ at our disposal. For instance, the field describing the first component becomes
\beq \label{ff}
\tilde \phi_1(\vecl) = \sum_{i} \epsilon^1_i \exp\lp - i \vecl\cdot \vecx_i\rp.
\enq
We assume that the measured ellipticities trace the shear fields, in the sense that the measured components are built out of the shear at that position plus some value unrelated to it,
\beq \begin{split} \label{rel}
\epsilon^1_i &= \gamma_1(\vecx_i) + \epsilon^1_{\textrm{int},\:i}
\\
\epsilon^2_i &= \gamma_2(\vecx_i) + \epsilon^2_{\textrm{int},\:i}.
\end{split}
\enq
The vector $\vecv$ relating the spectral matrices of the ellipticities and the convergence
is then obtained by plugging (\ref{ff}) with the above relations (\ref{rel}) in its definition (\ref{specmatrix}), and using the relation between shears and convergence in equation (\ref{distortion}). The result is 
\beq
\vecv =\bar n_\gamma  \bem \cos 2\phi_l \\ \sin 2\phi_l\enm.
\enq
where $\bar n_\gamma$ is the number density of galaxies for which ellipticity measurements are available.
Under our assumptions of uncorrelated intrinsic ellipticities,  with dispersions of equal magnitude $\sigma_\gamma^2$ for the two components, the noise matrix $N$ becomes 
\beq
N = \bar n_\gamma \bem \sigma_\gamma^2 & 0 \\ 0 & \sigma_\gamma^2\enm.
\enq
The effective noise, given in equation (\ref{noise}) is readily computed
\beq
N^\gamma_\eff = \frac{\sigma^2_\gamma}{\bar n_\gamma}.
\enq
\subsection{Sizes}
As noted in the main text, the apparent sizes of galaxies are modified by lensing, in the following way,
\beq
s^i_{\obs} = s^i_{\textrm{int}}\lp 1 + \alpha_s \kappa\rp,
\enq for some coefficient $\alpha_s$ which is unity in pure weak lensing theory. Denoting the number of galaxies for which sizes measurements are available by $n_s$, and the mean intrinsic size of the sample by $\bar s_{\textrm{int}}$,
the spectrum of the size field reduces, under the assumption of uncorrelated intrinsic sizes, to
\beq
C^s(l) = \bar n_s^2 \bar s_{\textrm{int}}^2 \alpha_s^2 C^{\kappa}(l) + \bar n_s \sigma_s^2.
\enq
The vector $\vecv$ and matrix $N$ are now numbers,  that are read out from the above equation, to be
\beq \begin{split}
v &= \bar n_s\bar s_{\textrm{int}} \alpha_s, \\
N &= \bar n_s \sigma_s^2.
\end{split}
\enq
leading to the effective noise
\beq
N_\eff^s = \frac{1}{\alpha_s^2}\lp \frac{\sigma_s}{\bar s_\textrm{int}} \rp^2\frac 1 {\bar n_s}
\enq
\subsection{Second order, flexion}
Denoting with $\bar n_{\mathcal F}$ and $\bar n_{\mathcal G}$ the number of galaxies for which $\mathcal F$ and $\mathcal G$ are measured, the vectors linking the flexion to convergence are
\beq
\vecv^{\mathcal F} = -il\bar n_{\mathcal F}\bem \cos \phi_l \\ \sin \phi_l\enm\enq
and
\beq
\vecv^{\mathcal G} = -il\bar n_{\mathcal G}\bem \cos 3\phi_l \\ \sin 3\phi_l\enm.
\enq
Using again the assumption of uncorrelated intrinsic components, we have the four dimensional diagonal noise matrix 
\beq
N = \bem \bar n_\mathcal F \sigma^2_\mathcal F \cdot 1_{2x2}& 0 \\0  & \bar n_\mathcal G \sigma^2_\mathcal G\cdot 1_{2x2}\enm,
\enq
leading to the effective noise, this time mode-dependent,
\beq 
\frac 1 { N^{\mathcal F \mathcal G}_\eff} = l^2\left(\frac{\bar n_F}{\sigma_F^2} + \frac{\bar n_G}{\sigma_G^2}\right).
\enq
\end{appendix}
\bibliographystyle{mn2e}
\bibliography{mybib}
\end{document}